\documentclass[showpacs,pre,aps,twocolumn,superscriptaddress]{revtex4-1}
\pdfoutput=1

\newcommand{\beq}{\begin{equation}}
\newcommand{\eeq}{\end{equation}}

\newcommand{\sigS}{\sigma_{\rm S}}
\newcommand{\sigB}{\sigma_{\rm B}}

\newcommand{\etaS}{\eta_{\rm S}^{\rm r}}
\newcommand{\etaSc}{\eta_{\rm S}^{{\rm r}*}}

\newcommand{\muB}{\mu_{\rm B}}
\newcommand{\muS}{\mu_{\rm S}}

\newcommand{\ind}{I}

\newcommand{\CC}{\mathcal{C}}

\usepackage{amsmath,amsfonts,amssymb,bm,graphicx,hyperref}
\usepackage{color}
\usepackage{array}

\begin{document}

\title{Critical point for de-mixing of binary hard spheres}
\author{Hideki Kobayashi}
\affiliation{Yusuf Hamied Department of Chemistry, University of Cambridge, Lensfield Road, Cambridge CB2 1EW, United Kingdom}

\author{Paul B. Rohrbach} 
\affiliation{DAMTP, University of Cambridge, Centre for Mathematical Sciences, Wilberforce Road, Cambridge CB3 0WA, United Kingdom}

\author{Robert Scheichl}
\affiliation{Institute for Applied Mathematics, Heidelberg University, INF 205, 69120 Heidelberg, Germany}
\affiliation{Department of Mathematical Sciences, University of Bath, Bath BA2 7AY, United Kingdom}

\author{Nigel B. Wilding}
\affiliation{H.H. Wills Physics Laboratory, University of Bristol, Royal Fort, Bristol BS8 1TL, United Kingdom}

\author{Robert L. Jack}
\affiliation{Yusuf Hamied Department of Chemistry, University of Cambridge, Lensfield Road, Cambridge CB2 1EW, United Kingdom}
\affiliation{DAMTP, University of Cambridge, Centre for Mathematical Sciences, Wilberforce Road, Cambridge CB3 0WA, United Kingdom}

\begin{abstract}
We use a two-level simulation method to analyse the critical point associated with demixing of binary hard sphere mixtures.  The method exploits an accurate coarse-grained model with two-body and three-body effective interactions.  Using this model within the two-level methodology allows computation of properties of the full (fine-grained) mixture.  The critical point is located by computing the probability distribution for the number of large particles in the grand canonical ensemble, and matching to the universal form for the $3d$ Ising universality class.  The results have a strong and unexpected dependence on the size ratio between large and small particles, which is related to three-body effective interactions, and the geometry of the underlying hard sphere packings.
\end{abstract}

\maketitle

Hard sphere systems are central to our understanding of many physical systems and phenomena, including the structure of the liquid state~\cite{hansen-book}; the behaviour of colloidal suspensions~\cite{Pusey1986,Auer2001,Royall2013-hard};  jamming and glass transitions~\cite{Weeks2000,Parisi2010}; and packing problems~\cite{Graaf2011,vanAnders2014}.  In equilibrium statistical mechanics, hard-particle systems are simple and elegant, because every allowed configuration has the same  statistical weight.  Despite this simplicity, these systems are of practical importance: they are amenable to experiments~\cite{Pusey1986,Weeks2000,Sacanna2010,Royall2013-hard};  and they support complex behaviour including a variety of phase transitions~\cite{Alder1957,Wood1957,Bernard2011,Damasceno2012,Gantapara2013,Ashton2015}, which continue to challenge theoretical and computer simulation methods.
%This article considers 
We focus here on \emph{mixtures of large and small hard spheres}, which are predicted to undergo fluid-fluid phase separation (de-mixing), if the size disparity and the concentrations are large enough~\cite{Biben1991,Dijkstra1998,Dijkstra1999-pre,RED}.   The phase where the large particles predominate corresponds to a (metastable) colloidal liquid~\cite{Poon2002}.  Contrary to the usual intuition that liquids are stabilised by attractive forces, this phase appears in an equilibrium system with additive mixing rules and without any attractive forces between particles.  This illustrates the depletion mechanism for de-mixing~\cite{Asakura1954,Lekkerkerker1992,Poon2002}, which is one of the prototypical mechanisms for fluid-fluid phase separation.

Given its status as a theoretical benchmark, it may be surprising that this fluid-fluid phase separation of hard spheres has never been accurately characterized.  Buhot and Krauth~\cite{Krauth1998} showed that large particles cluster together strongly, in small systems at moderate overall volume fractions; Dijkstra, van Roij and Evans~\cite{Dijkstra1999-prl} analysed fluid-solid and solid-solid de-mixing.
%, for which their theory is accurate~\cite{Dijkstra1998,Dijkstra1999-pre}.  
These numerical studies confirm that depletion leads to strong effective interactions in these systems~\cite{Krauth1998}, whose behaviour is captured semi-quantitatively by theoretical arguments~\cite{Biben1991,Dijkstra1998,Dijkstra1999-pre}.
However, the critical point for de-mixing has never been observed directly, nor have the coexisting fluid phases.

\begin{figure}[b]
%\vspace{4cm}
\includegraphics[width=8cm]{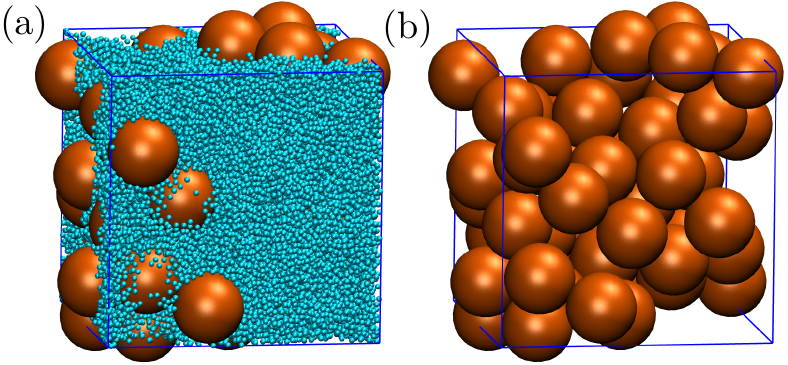}
\caption{(a) Snapshot of a binary hard sphere system at size ratio 11:1, near criticality ($\etaS=0.302$).  The box size is $\tilde L=44$, there are $N=50$ large particles.  (b) The same configuration with the small particles removed: this is a configuration of the CG model. 
}
\label{fig:snap}
\end{figure}

The reason for this state of affairs
is that de-mixing involves collective behaviour of the large particles, which can only be observed if their number is great enough.  Additionally, a depletion effect that is strong enough to produce demixing requires a large disparity in size between the particles, and a large concentration of the small particles.   Hence one must analyse configurations with very many small particles, and there is also a huge disparity between the time scales on which the two species relax.  Fig.~\ref{fig:snap} illustrates the severe challenges that this poses for computer simulation: the systems are extremely crowded, and they include many particles of disparate sizes, with significant interparticle correlations.  This complexity also means that exact theoretical computations are out of reach, so efficient numerical methods are necessary  for accurate results.

This work uses a two-level numerical method~\cite{Kobayashi2019} to characterise the critical point for fluid-fluid phase separation in hard sphere mixtures.
The first level of the method relies on an accurate coarse-grained (CG) model where the small particles are integrated out, providing an effective theory for the large ones.   Then, the second level restores the small particles, providing (numerically) exact results for the full mixture.
The method was previously validated for the Asakura-Oosawa model~\cite{Asakura1954}, which is a much simpler example of de-mixing, for which an exact CG model is available.
The results presented here show that the method is viable in complex systems, finally allowing direct observation of the phase transition in the hard sphere system.  The results also reveal new physics, in that the packing of the hard particles influences the phase transition via three-body depletion interactions, which have been neglected in previous theories~\cite{Biben1991,Dijkstra1998,Dijkstra1999-pre,RED,Largo2006,Amokrane2006}.  As such, our results confirm the qualitative picture proposed in~\cite{Biben1991,Dijkstra1998,Dijkstra1999-pre}, so that arguments against fluid-fluid demixing are not correct~\cite{Lopez2013,Santos2020}.   But they also highlight that the standard two-body depletion theories are not adequate for accurate characterisation of this important phase transition~\cite{Ashton2011depletion}.

We analyse a binary mixture of hard spheres whose diameters are $\sigS$ (small particles) and $\sigB = \ell \sigS$ (big particles, so $\ell>1$).  We use a cubic simulation box of linear size $L$ with periodic boundaries, in the grand canonical ensemble.
%Since the particles are hard, we work in units where $k_{\rm B}T=1$, without loss of generality. 
The relevant dimensionless parameters are the size ratio $\ell$, the system size $\tilde{L}=L/\sigS$ and the chemical potentials $\mu_{\rm B},\mu_{\rm S}$ (measured relative to $k_{\rm B}T$).  We parameterise $\mu_{\rm S}$ in terms of the (reservoir) small-particle volume fraction $\etaS$, using an accurate equation of state~\cite{Kolafa2004}, see Appendix~\ref{sec:theory}.
A configuration of the system has $N$ large particles with positions $\bm{R}_1,\dots,\bm{R}_N$ and $n$ small particles
with positions $\bm{r}_1,\dots,\bm{r}_n$.

\begin{figure*}
\includegraphics[width=18cm]{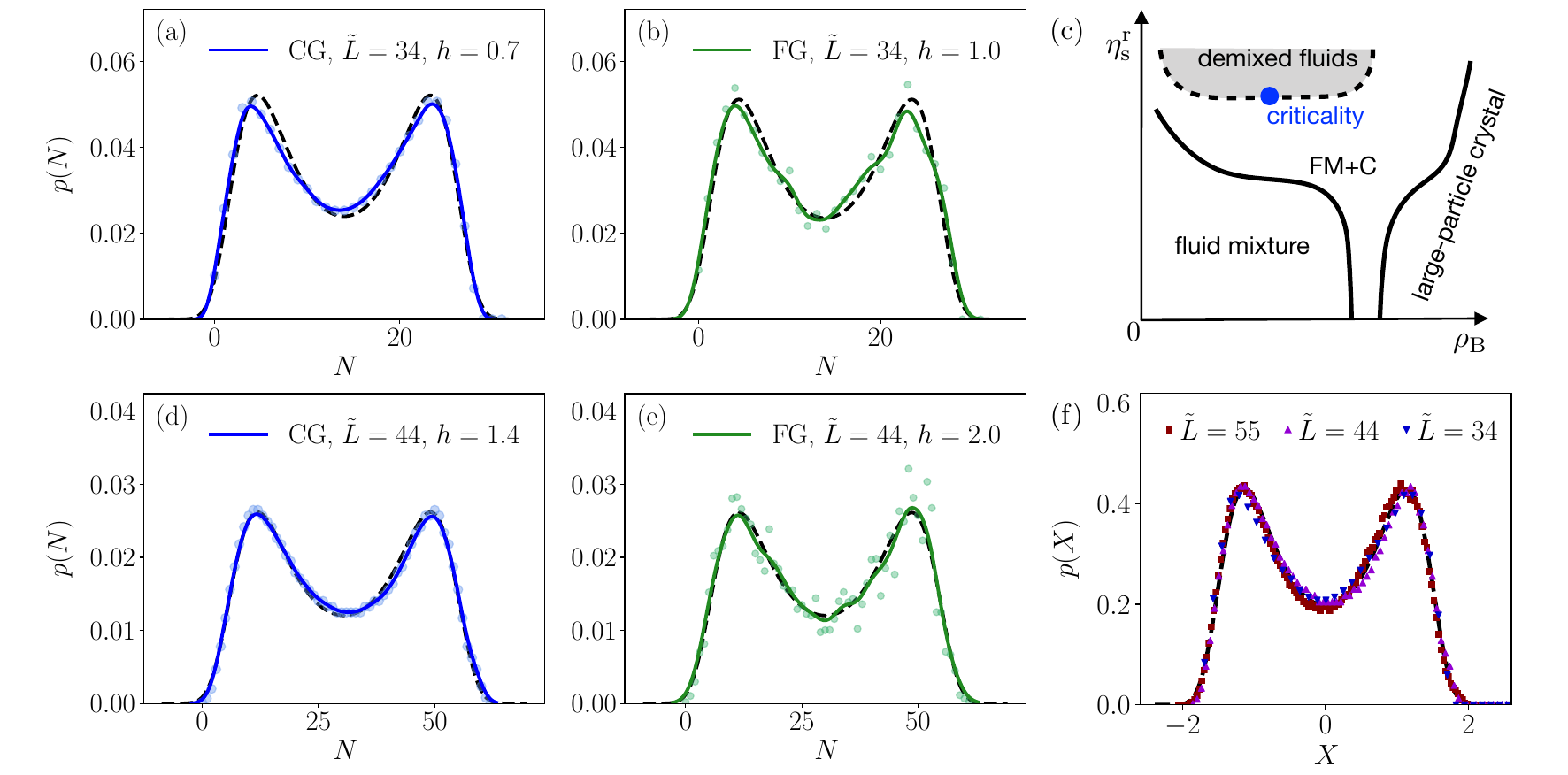}
\caption{%
Results for $\ell=11$.  Histograms for the number of large particles for CG model \textbf{(a,d)} and FG models \textbf{(b,e)},
with $\etaS=0.3020,0.3010$ for $\tilde{L}=34,44$, respectively.
The universal critical form~\cite{Tsypin2000} is shown by dashed lines.  The underlying data are shown as points and the solid lines are Gaussian kernel density estimates with width parameters $h$ as shown.  
\textbf{(c)}~Schematic phase diagram, following~\cite{Dijkstra1999-pre}, as a function of $\rho_{\rm B}=N/L^3$ and $\etaS$.  The critical point for de-mixing is indicated, together with the coexistence region (shaded).  The demixed state is metastable with respect to crystallization of the large particles (the coexistence region between a fluid mixture and a crystal is labelled as FM+C).  The region of coexistence between two crystal phases is omitted for simplicity, see~\cite{Dijkstra1999-pre}.
\textbf{(f)}~Finite-size scaling collapse for CG data at $\tilde{L}=34,44,55$, results for $\tilde{L}=55$ have $\etaS=0.3006$.
}
\label{fig:histos}
\end{figure*}

Fig.~\ref{fig:histos}(c) illustrates the phase diagram proposed in~\cite{Dijkstra1998,Dijkstra1999-pre}.
As a signature of demixing, we seek the critical point.
Define $p(N)$ as the probability that the system contains $N$ large particles.  For large systems in a single-phase regime, $p(N)$ is unimodal and Gaussian.
As one approaches the critical point $(\muB^*,\etaSc)$,%_{\rm S}^*)$, 
the distribution $p(N)$ broadens; at the critical point the large particles form a fractal structure, and $p(N)$ has a characteristic (universal) scaling form~\cite{wilding1995critical,Tsypin2000}.  
For $\etaS > \etaSc$ %_{\rm S}^*$ 
one expects a phase coexistence line in the $(\muB,\etaS)$ plane, where $p(N)$ is bimodal.

We locate the critical point
by matching the observed $p(N)$ to its universal scaling form~\cite{wilding1995critical,Bruce2003}, corresponding to the $3d$ Ising universality class.  
Since $N$ (or equivalently the concentration $N/\tilde{L}^3$) is the natural order parameter for the de-mixing transition, it is natural to work in the grand canonical ensemble, where the finite-size scaling of the critical fluctuations is well understood~\cite{wilding1995critical}.   

Sampling $p(N)$  is not tractable by standard methods -- it requires that large particles are inserted and removed from the system, which is almost impossible in crowded environments like that shown in Fig.~\ref{fig:snap}.
This problem is avoided by the two-level method.  We outline the approach, see~\cite{Kobayashi2019} and Appendix~\ref{sec:theory}
%Supplementary Material (SM)~\cite{SM} 
for details.
The critical points of interest generically occur for parameters where the fluid phase is metastable with respect to crystallisation of the large particles~\cite{Dijkstra1999-pre} (recall Fig.~\ref{fig:histos}c), in fact the two-level method helps to control for crystallisation (see Appendix~\ref{sec:supp-results}).
%\cite{SM}.

The method relies on a coarse-grained (CG) model, where the small particles are integrated out, leaving only the large ones.  
It involves an effective interaction among the large particles, the corresponding energy is
\begin{multline}
E_{\rm CG}({\cal C}) =  N \Delta \mu +  \sum_{1\leq i<j \leq N} V_2(R_{ij}) 
\\ + \sum_{1 \leq i < j < k \leq N } V_3(R_{ij},R_{ik},R_{jk})
\label{equ:cgPot}
\end{multline}
where $R_{ij}=|\bm{R}_i-\bm{R}_j|$ is the distance between particles $i$ and $j$; also $V_2$ and $V_3$ are two- and three-body effective interactions, and the term $N\Delta \mu$ ensures that the chemical potentials coincide between FG and CG models.  The $\Delta\mu,V_2,V_3$ are obtained by grand-canonical Monte Carlo  (GCMC) simulation of small particles in systems which contain a few fixed large particles, see~\cite{Kobayashi2019} and Appendix~\ref{sec:comp}.

The resulting CG model is highly accurate but it is not a perfect description of the large-particle behavior.  Hence the second step of the method, which computes  the difference between the CG result and the result for the full (fine grained, FG) model.   
Recalling that $p(N)$ is the probability that the FG model has $N$ large particles, define $p_{\rm CG}(N)$ as the corresponding quantity for the CG model.  Then
\beq
p(N) = p_{\rm CG}(N) + \Delta p(N)
\label{equ:pN-FC}
\eeq
where $\Delta p(N)$ is the coarse-graining error.

The distribution $p_{\rm CG}$ is computed by GCMC simulation of the coarse model and
 the correction $\Delta p$ is calculated following~\cite{Kobayashi2019}, using a free energy estimate based on Jarzynski's equality~\cite{Jarzynski1997,Crooks2000,Neal2001,Hummer2001}.
The computation of $\Delta p(N)$ distinguishes our approach from traditional coarse-graining methods~\cite{Noid2008,Praprotnik2007,Ouldridge2011,Mladek2013,Pak2018} in which the main concern is that the CG model is as accurate as possible, but its error is not usually quantified. 
In practice, our CG model is accurate enough that the correction $\Delta p$ will turn out to be small.  
(Computation of this correction has similarities with free energy perturbation theory~\cite{Zwanzig1954}, as recently exploited to correct coarse-graining errors for machine-learned potentials~\cite{Cheng2019}.)

Fig.~\ref{fig:histos} shows results for $\ell=11$.  For a tractable analysis, we considered relatively small system sizes $\tilde L=34,44$, which are between 3 and 4 times the diameter of a large particle.  The behavior of $p_{\rm CG}(N)$ is shown in Fig.~\ref{fig:histos}(a,d).  By adjusting $\etaS$ and $\mu_{\rm B}$, we obtained estimates of the critical point, where the distribution $p_{\rm CG}$ matches its universal critical form (black dashed line), which has been scaled to give the correct mean and variance.  The systems are small but the fit to the universal form is good.  The agreement with the universal distribution ensures that cumulant ratios~\cite{binder-book} 
are also in agreement with their
 universal values at criticality.

Turning to the FG model, we estimate the correction $\Delta p$ , and hence the distribution $p(N)$ for the binary mixture.  
%For the system with $\tilde L=44$, we considered $M=1280$ coarse configurations. 
The method requires $M$ configurations of the CG model which we denote as $\CC_1,\CC_2,\dots,\CC_M$, obtained by GCMC simulation.  (Specifically, we take $M=1280$.)  For each coarse configuration, we then perform a GCMC simulation for the small particles, with the large ones held fixed.  This yields a reweighting factor $\hat\omega_\alpha$ (see Appendix~\ref{sec:theory}) then
$\Delta p(N)$ is estimated as
\beq
\Delta\hat{p}(N) =  
  \sum_{\alpha=1}^M  \left( \hat\omega_\alpha -1 \right) \ind_N({\cal C_\alpha})
  \label{equ:hat-dp}
\eeq
where $\ind_N({\cal C_\alpha})=1$ if $\CC_\alpha$ contains $N$ large particles, and $\ind_N({\cal C_\alpha})=0$ otherwise.

Results for $p(N)$ are shown in Fig.~\ref{fig:histos}(b,e), including individual estimates of $p(N)$, and (smoothed) kernel density estimates of $p$, based on the same data.  The resulting distributions match the universal scaling form, indicating that the FG model is indeed very close to its critical point, see also~\cite{wilding1995critical,Debenedetti2020}. 
For a finite-size scaling analysis, we recenter and scale the particle number $N$ to zero mean and unit variance:
\beq
X = \frac{ N - \langle N \rangle}{ \Delta_N }, \qquad \Delta_N = \sqrt{\langle N^2\rangle - \langle N \rangle^2} \; .
\label{equ:def-X}
\eeq
Fig.~\ref{fig:histos}(f) shows additional finite-size scaling results for the CG model at $\ell=11$, including results at a larger system size $\tilde{L}=55$.  
These results are consistent with behavior in the Ising universality class, although the systems are small enough that corrections to scaling are significant, see \cite{wilding1995critical} and Appendix~\ref{sec:supp-results}.

It can be shown that the estimates of $p(N)$ are asymptotically unbiased~\cite{Kobayashi2019}, but they do suffer from large variance 
if either (i) the CG model is not sufficiently accurate or (ii) the free-energy computations are performed too quickly~\cite{Oberhofer2009}.
These effects can lead to fat-tailed distributions of reweighting factors $\hat\omega_\alpha$, so that the estimate $\Delta\hat{p}$ starts to be dominated by a few (non-typical) configurations $\CC_\alpha$.  This can be easily checked from the numerical data, providing a consistency check on the method.  
 In fact efficient performance with moderate $M$ (as used here) requires a typical coarse-graining error significantly less than $k_{\rm B}T$ in the total energy $E_{\rm CG}$.
The behaviour of the weights $\hat\omega_\alpha$ is discussed in Appendix~\ref{sec:supp-results}
%SM~\cite{SM}, 
showing that this condition holds. 
We also note the FG data points in Fig.~\ref{fig:histos} are scattered around the kernel density estimate, this indicates the size of the numerical errors (which would be very large variance if the CG model was not accurate)

\begin{figure}
\includegraphics[width=8.5cm]{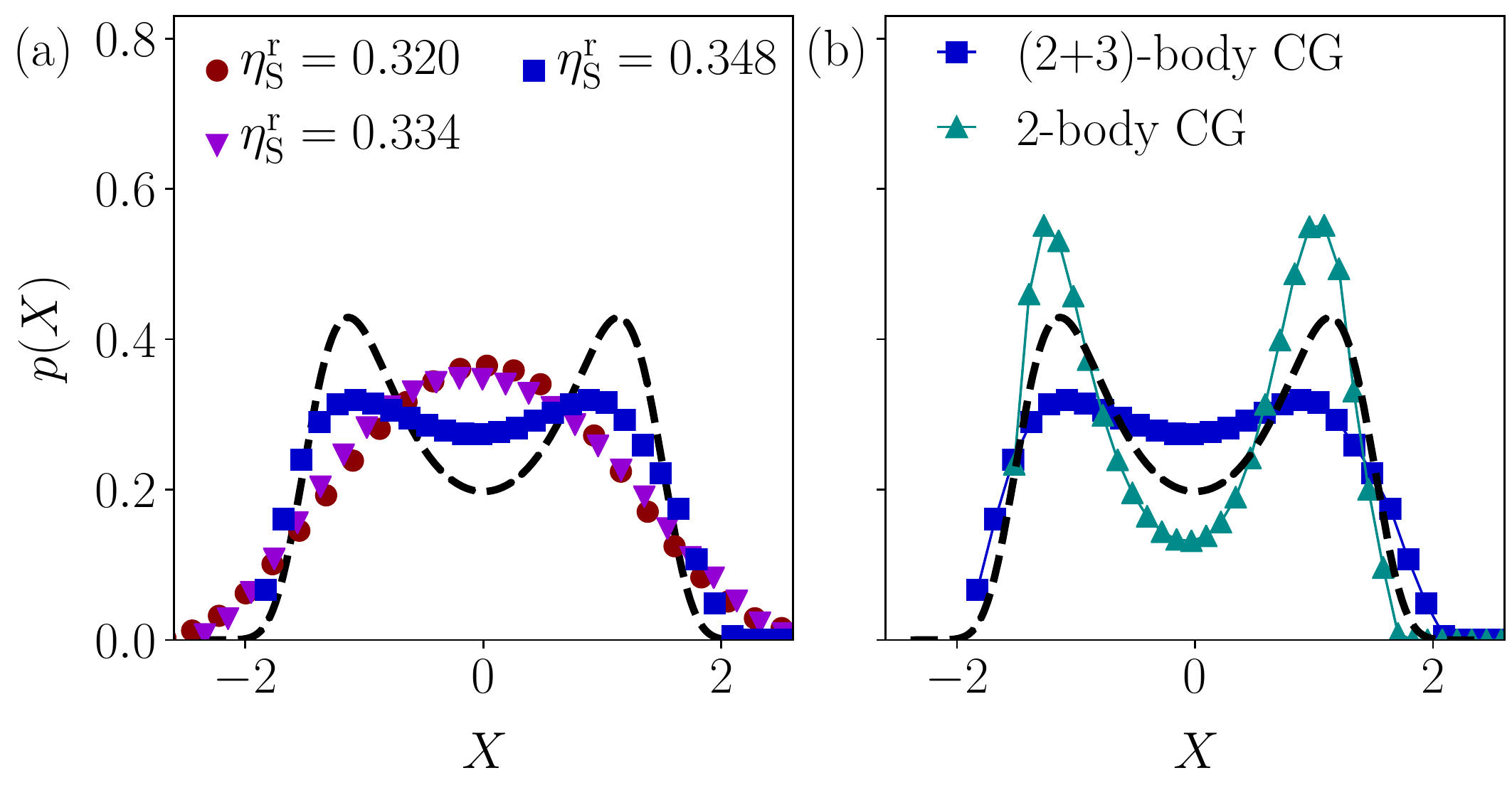}
\caption{Distributions of the order parameter $X$ in the CG model at $\ell=10$.
The system size is $L=3\sigB$ (so $\tilde L = 30$), dashed lines indicate the universal critical form. (a)~Results for increasing $\etaS$, indicating that $\eta_{\rm S}^*\gtrsim 0.348$ (the results at this largest $\etaS$ appear close to criticality, but demixing has not yet occurred).
(b)~Comparison of the CG model (\ref{equ:cgPot}) [labelled as $(2+3)$-body] and a 2-body CG model ($V_3=0$), both at $\etaS=0.348$.  The three-body interaction suppresses de-mixing. }
\label{fig:histos-sig10}
\end{figure}

In contrast to the results for $\ell=11$, the behavior of the CG system for $\ell=10$ is shown in Fig.~\ref{fig:histos-sig10}(a), for a small system $L=3\sigB$.  
The distributions of $X$ in Fig.~\ref{fig:histos-sig10}(a) are ``less bimodal'' than the (universal) critical form, indicating that if this system has a critical point, it has $\etaSc\gtrsim 0.35$.  For such high volume fractions, any computations involving small particles become challenging, including accurate estimation of the CG potential, so we have not explored further into this regime.
In the range shown, 
the  three-body effective interactions for $\ell=10$ are repulsive, especially for larger $\etaS$, see Appendix~\ref{sec:supp-results}.  To illustrate their effect, Fig.~\ref{fig:histos-sig10}(b) compares the CG model with a similar ($2$-body CG) model without any three-body interactions ($V_3=0$).  For the two-body CG model, it is clear that $\etaSc<0.348$, but the three-body interaction drives the critical point to larger $\etaS$.  
%We were not able to locate accurately the corresponding critical point: the three-body potential tends to suppress de-mixing but this effect is hard to calculate, because of the high computational cost of accurate CG potentials at such large $\etaS$. 

To summarize: Fig.~\ref{fig:histos} demonstrates a de-mixing critical point in CG and FG models of binary hard spheres with $\ell=11$ and $\etaSc\approx 0.30$, but
 Fig.~\ref{fig:histos-sig10} shows that for $\ell=10$ the corresponding critical point is beyond the reach of our numerics, $\etaSc\gtrsim0.35$.  For $\ell=10$, previous estimates of $\etaSc$~\cite{Biben1991,Dijkstra1998,Dijkstra1999-pre,RED,Largo2006} were smaller ($0.29$--$0.32$), but such treatments assumed that 2-body CG models are accurate.  Fig.~\ref{fig:histos-sig10}(b) shows explicitly that three-body effective interactions suppress de-mixing at $\ell=10$, explaining the difference in $\etaSc$.  By contrast, for $\ell=11$ the two-body CG model is more accurate; indeed the Noro-Frenkel criterion~\cite{Noro2000} holds quite accurately at the critical point (see Appendix~\ref{sec:supp-results}).
 
 \newcommand{\phin}{\phi_{\rm in}}
 
 \begin{figure}
\includegraphics[width=8.5cm]{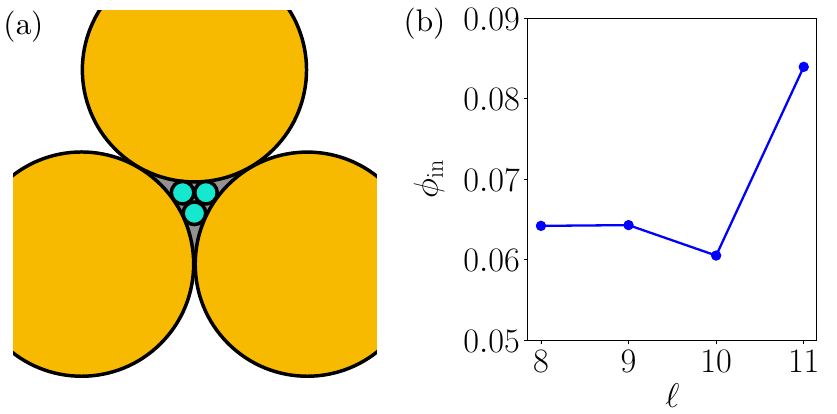}
\caption{Packing of large and small particles.  (a)~Planar configuration with three large particles touching each other and three small particles in the (grey shaded) space between them.
(b)~Measure of packing efficiency $\phi_{\rm in}$ as defined in main text, for $\etaS=0.32$.  This quantity increases sharply between $\ell=10$ and $\ell=11$.}
\label{fig:packing}
\end{figure}
 
For a physical explanation of these substantial differences between $\ell=10,11$,
%For a physical explanation of the significant differences between $\ell=10,11$ 
note that de-mixing is favoured if the colloidal liquid (large-$N$) phase supports efficient packing.  The depletion effect makes it likely that large particles are very close to each other, so it matters whether the small particles pack efficiently into the gaps between them.  Planar configurations similar to Fig.~\ref{fig:packing}(a) are efficient for packing, but it can be shown that they are only possible for $\ell\geq5+\sqrt{24}\approx9.9$.   Hence, such configurations are very rare for $\ell=10$ (which is close to the marginal case), but they are much more common for $\ell=11$.  
To show this explicitly, we used GCMC simulation for small particles to compute a (normalised) number density $\phin$ in the shaded grey region of Fig.~\ref{fig:packing}(a), which we interpret as a measure of packing efficiency, see Appendix~\ref{sec:supp-results}.  For the representative volume fraction $\etaS=0.32$,  Fig.~\ref{fig:packing}(b) shows that the packing efficiency $\phin$ increases sharply between $\ell=10$ and $\ell=11$, which explains the enhanced de-mixing in the latter case.  (Other signatures of more efficient packing at $\ell=11$ are shown in Appendix~\ref{sec:supp-results}
%~\cite{SM}, 
including a discussion of three-body effective interactions.)  

We note that two-body CG models are widely used in soft matter~\cite{LIKOS2001267}, and such models are generally expected to be accurate in hard sphere mixtures with large $\ell$~\cite{Dijkstra1998,Dijkstra1999-pre}.
Since three-body interactions turn out to be relevant even in this case, our results indicate that caution is advisable when applying two-body CG models in soft matter~\cite{Ashton2011depletion}.

We close with a few comments on the two-level method~\cite{Kobayashi2019}.  To characterise the critical point to high accuracy, we use GCMC simulation and match $p(N)$ to its critical form.  In this setting, the two-level method sidesteps the problem of inserting large particles into the crowded environment shown in Fig.~\ref{fig:snap}, because particle insertion is performed at the CG level, while the small particles only appear in the second (FG) level.  The method requires a very accurate CG model~\cite{Kobayashi2019} and considerable computational effort, but this is mitigated by the fact that the FG level is trivial to parallelise.   This method falls into the general class of multi-level approaches~\cite{Giles2008,Anderson2012,Hoang2013,Dodwell2015,Beskos2017,Dodwell2019};
the results presented here 
are further evidence that multi-level  coarse-graining methods have useful application in the 
physical sciences~\cite{Kobayashi2019,Rosin2014,Jansen2020,Lovbak2020} (see also~\cite{Brandt2001,Brandt2002,Brandt2003-chapter}), especially if it can be combined
 with machine-learned effective potentials~\cite{Behler2007,Bartok2010,Schutt2018,Gartner2020}, similar to~\cite{Cheng2019}.
We look forward to future work in this direction.

%\begin{acknowledgments}
We thank Daan Frenkel and Bob Evans for helpful discussions.
This project was supported by the Leverhulme Trust (grant RPG-2017-203).  RLJ and HK are also grateful to the EPSRC for support in the later part of the project (grant EP/T031247/1).
%\end{acknowledgments}

\newcommand{\EqECG}{1}
\newcommand{\EqHatDp}{3}

\newcommand{\FigHistos}{2}
\newcommand{\FigHistosSigTen}{3}
\newcommand{\FigPhi}{4}

\newcommand{\RefDijkstraPRE}{18}
\newcommand{\RefRED}{19}
\newcommand{\RefKoba}{25}
\newcommand{\RefAshton}{23}
\newcommand{\RefKolafa}{30}
\newcommand{\RefWild}{33}
\newcommand{\RefJarz}{35}
\newcommand{\RefCrooks}{36}
\newcommand{\RefZwanzig}{44}
\newcommand{\RefNoro}{49}
\newcommand{\RefMiller}{\cite{Miller2003}}
\newcommand{\RefVega}{\cite{Vega1998}}
\newcommand{\RefStell}{\cite{Stell1991}}
\newcommand{\RefBaxter}{\cite{Baxter1968}}

\begin{appendix}

%This Supplementary Information contains three main sections.  Sec.~\ref{sec:theory} reviews the theory of the two-level method~[\RefKoba], as it applies to the binary hard sphere system.  
%Sec.~\ref{sec:comp} provides methodological details on the computation of the CG potentials and the simulations that are used to analyse the FG model.
%Sec.~\ref{sec:supp-results} provides additional results and discussion to support the conclusions of the main text.
%
%The cited references in this document are listed in the main text.

\section{Theory}
\label{sec:theory} 

\subsection{FG model (binary mixture)}
\label{sec:fg}

We define the relevant properties of the binary hard sphere system (BHS) and the corresponding CG model.
Following~[\RefKoba], we denote the large-particle (coarse) degrees of freedom by
\beq
{\cal C} = (N,\bm{R}_1,\dots,\bm{R}_N) \; ,
\label{equ:CC}
\eeq
and the small-particle (fine) degrees of freedom by
\beq
{\cal F} = (n,\bm{r}_1,\dots,\bm{r}_n) \; .
\label{equ:FF}
\eeq
Define a function
$
 e_{\rm BHS}({\cal C},{\cal F}) 
$
such that $e_{\rm BHS}=1$ if none of the hard spheres overlap each other, and $e_{\rm BHS}=0$ otherwise.
Then the Boltzmann weight for any configuration of the BHS system is
\beq
w_{\rm BHS}({\cal C},{\cal F}) = e_{\rm BHS}({\cal C},{\cal F}) \frac{ \exp(\muB N + \muS n)  }{  N! \, n! }
\label{equ:w-BHS}
\eeq
and the probability density for configurations in the grand canonical ensemble is 
\beq
p_{\rm BHS}({\cal C},{\cal F})=\frac{ w_{\rm BHS}({\cal C},{\cal F}) }{ \sigB^{3N}\sigS^{3n} \Xi} \; ,
\label{equ:p-bhs}
\eeq 
where the normalization constant $\Xi$ is the grand-canonical partition function.  Specifically
\beq
\Xi = \sum_{N,n}  \int d\bm{R}_1\dots d\bm{R}_N \,  d\bm{r}_1\dots d\bm{r}_n \frac{ w_{\rm BHS}({\cal C},{\cal F}) }{ \sigB^{3N}\sigS^{3n}} 
\eeq
where each particle position is integrated over the simulation box (which is a cube of size $L$).  Hence $\Xi$ depends on $\ell,\muB,\muS,\tilde L$.

Averages in the FG/CG models are denoted by $\langle \cdot \rangle_{{\rm FG}/{\rm CG}}$.  
Specifically, if $A$ is an observable quantity in the FG model then
\beq
\langle A(\CC,{\cal F}) \rangle_{\rm FG} = \sum_{N,n} \int d\bm{R}^N \, d\bm{r}^n\, A(\CC,{\cal F}) p_{\rm BHS}({\cal C},{\cal F})
\label{equ:ave-fg}
\eeq
where the integrals are over all particle positions, within the simulation box.

Here and in the following, note that weight functions like $w_{\rm BHS}$ are dimensionless (and not normalised as probability distributions), but $p$ indicates a normalised probability density.

It is natural to define the small-particle volume fraction as 
\beq
\eta_{\rm S} =  \frac{\pi}{6\tilde{L}^3}\langle n \rangle_{\rm FG}.
\eeq  
The reservoir volume fraction $\etaS$ is the value of $\eta_{\rm S}$ that one obtains in a system with no large particles at all, as $\tilde{L}\to\infty$.  This only depends on $\muS$ and can be estimated very accurately using the equation of state of (\RefKolafa).  This $\etaS$ depends monotonically on $\muS$; it is used to parameterise the dependence of the results on $\muS$ (whose value is not particularly intuitive) in terms of the more natural parameter $\etaS$.

\subsection{CG model}
\label{sec:cg}

 The
coarse degrees of freedom $\cal C$ from \eqref{equ:CC} describe configurations of the CG model.
Define $e_{\rm HS}({\cal C})=1$ if none of the large particles overlap with each other and $e_{\rm HS}({\cal C})=0$ otherwise, analogous 
to $e_{\rm BHS}$ above.
The Boltzmann weight for the CG model is
\beq
w_{\rm CG}({\cal C}) = e_{\rm HS}({\cal C}) \frac{ \exp[ \mu_B N - E_{\rm CG}({\cal C}) ]}{ N! }
\label{equ:wCG}
\eeq
where the effective interaction energy $E_{\rm CG}$ is given in Eq.~(\EqECG) of the main text.
Similar to the FG case define 
\begin{align}
p_{\rm CG}({\cal C})&=\frac{ w_{\rm CG}({\cal C}) }{ \sigB^{3N} \Xi_{\rm CG}} \; ,% \qquad
\nonumber\\
\Xi_{\rm CG} &= \sum_{N}\int  d\bm{R}_1\dots d\bm{R}_N  \frac{ w_{\rm CG}({\cal C}) }{ \sigB^{3N} }  \; .
\end{align}
If $A$ is an observable quantity in the CG model then its average is
\beq
\langle A(\CC) \rangle_{\rm CG} = \sum_{N} \int  d\bm{R}_1\dots d\bm{R}_N A(\CC) p_{\rm CG}({\cal C})
\label{equ:ave-cg}
\eeq

\subsection{Transformation between models, and computation of $\Delta p$}
\label{sec:level2}

To connect the CG and FG models, we (formally)
integrate out the small particles from the FG model.  The result is an effective Boltzmann weight for the large particles alone, which is
\beq
w_{\rm eff}({\cal C}) = %\frac{ \exp(\mu_B N) }{ N! } 
\sum_{n=0}^\infty \int {\rm d}\bm{r}_1 \dots {\rm d}\bm{r}_n \, \frac{ w_{\rm BHS}({\cal C},{\cal F}) }{ \sigS^{3n} } \; .
\label{equ:weff-integral}
\eeq
Now define $\Phi({\cal C})$ as the grand-canonical free energy of the small particles, evaluated for a fixed large-particle configuration ${\cal C}$:
\beq
\Phi({\cal C}) = - \log \sum_{n=0}^\infty \int {\rm d}\bm{r}_1 \dots {\rm d}\bm{r}_n \, \frac{ e_{\rm BHS}({\cal C},{\cal F}) \exp(\muS n) }{ \sigS^{3n} \, n! } \; .
\label{equ:def-Phi}
\eeq
[This quantity is finite as long as the large particles do not overlap, $e_{\rm HS}(\CC)=1$.  If $e_{\rm HS}(\CC)=0$ then $e_{\rm BHS}(\CC,{\cal F})=0$ also, so $w_{\rm eff}(\CC)=0$.]
Comparing the integrals in the two preceding equations and using (\ref{equ:w-BHS}), we find
\begin{equation}
w_{\rm eff}({\cal C})  = e_{\rm HS}({\cal C}) \frac{ \exp[\muB N  -\Phi({\cal C})] }{ N! } \; .
\label{equ:weff-Xi}
\end{equation}

A perfect CG model would have $w_{\rm CG}({\cal C}) = w_{\rm eff}({\cal C}) / \Xi_0$ for some constant $\Xi_0$ (independent of ${\cal C}$): in this case the CG model would exactly reproduce the behavior of the large particles in the FG model.  Comparing \eqref{equ:wCG} with \eqref{equ:weff-Xi}, this amounts to $E_{\rm CG}({\cal C}) = \Phi({\cal C}) + \log \Xi_0$.  However, in the absence of an exact coarse-graining computation, such a perfect CG model is not available.  

Still, one can make progress if the CG model provides a good approximation to $w_{\rm eff}$, because averages in the FG and CG models are related.  Let $A$ be an observable quantity that depends only on the large particles.  Combining the ingredients gathered above one finds
\begin{align}
\langle A({\cal C}) \rangle_{\rm FG} &= \frac{1}{Z} \left\langle A({\cal C}) \frac{w_{\rm eff}({\cal C})}{w_{\rm CG}({\cal C}) } \right\rangle_{\rm CG}  \; ,
%\qquad 
\nonumber\\
Z&= \left\langle \frac{ w_{\rm eff}({\cal C}) }{ w_{\rm CG}({\cal C}) }  \right\rangle_{\rm CG}  \; .
\label{equ:two-level}
\end{align}

Now define $\ind_N({\cal C})$ to be equal to unity 
if the system contains $N$ large particles and zero otherwise.
Hence $p(N) = \langle \ind_N({\cal C}) \rangle_{\rm FG}$ so using Eq.~\EqHatDp\ with
 Eqs.~\ref{equ:wCG},\ref{equ:weff-Xi},\ref{equ:two-level} yields
\beq
\Delta p(N) =  \left\langle \ind_N({\cal C}) \left[ \frac{ W({\cal C})}{Z} -1 \right] \right\rangle_{\rm CG} \; .
\label{equ:dp-rwt}
\eeq
with 
\beq
W(\CC) = \exp\left[E_{\rm CG}({\cal C}) - \Phi({\cal C})\right]
\label{equ:def-W}
\eeq
[Similarly, one may write $Z=\langle W(\CC)  \rangle_{\rm CG}$.]
This means that if $\Phi$ can be computed (or estimated) then so can $\Delta p$, and hence also $p$.  Moreover, \eqref{equ:dp-rwt} is an average in the CG model, which is computationally tractable.  The same idea is used in free-energy perturbation theory~[\RefZwanzig], to relate complicated models to simpler (more tractable) ones.

\subsection{Estimation of small-particle free energy $\Phi$}
\label{app:Phi}

To make use of (\ref{equ:dp-rwt}) in practice, we require a computational estimate of $W(\CC)$.  The object $\Delta \hat{p}$ in (\EqHatDp) is an estimator  for (\ref{equ:dp-rwt}), with $\hat\omega_\alpha$ in (\EqHatDp) corresponding to the ratio $W(\CC)/Z$ in (\ref{equ:dp-rwt}).
We estimate ${\rm e}^{-\Phi(\CC)}$ using a method based on Jarzynski's equality~[\RefJarz], as described in~[\RefKoba].  We give a short outline here.
It is important that $\Phi({\cal C})$ depends on the small-particle chemical potential $\muS$, via $w_{\rm BHS}$.
First select a very small chemical potential $\muS=\mu_0$, in which case the integral can be estimated directly from a grand canonical simulation.  Denote the corresponding value of $\Phi(\CC)$ by $\Phi_0(\CC)$.  Then, starting from an equilibrated system at chemical potential $\mu_0$, perform an GCMC simulation during which the small particle chemical potential increases in $K$ steps from $\mu_0$ to $\mu_{\rm S}$.
Then compute
\beq
{\cal I}(\CC) = \sum_{j=1}^K n_j \Delta \mu_j 
\label{equ:def-I-step}
\eeq
where
$\Delta \mu_j$ is the change in $\mu$ on the $j$th step and $n_j$ is the number of small particles in the system when that step takes place.
Since this quantity is the work done to insert the small particles, it follows from Crooks' theorem~[\RefCrooks] that
$
{\rm e}^{{\cal I}(\CC) - \Phi_0(\CC)} 
$
is an unbiased estimate of ${\rm e}^{-\Phi(\CC)}$.  That is, 
\beq
\left\langle {\rm e}^{{\cal I}(\CC) - \Phi_0(\CC)} \right\rangle_{\rm MC} = {\rm e}^{-\Phi(\CC)}
\label{equ:crooks}
\eeq 
where the average is over many realisations of the random MC algorithm (always with the same large particle configuration $\CC$). Hence 
\beq
\hat{W}(\CC) = {\rm e}^{E_{\rm CG}(\CC) + {\cal I}(\CC) - \Phi_0(\CC) }  
\label{equ:hat-W}
\eeq
is an unbiased estimate of $W(\CC)$. % defined in Eq.~\EqDefW. % (\ref{equ:def-W}).
Note that this result does not depend on the parameters of the GCMC simulation that was used to compute ${\cal I}$.  However, the variance of the estimate $\hat{W}$ does depend strongly on these parameters, which must be chosen judiciously for the method to be effective. 
%(see Appendix~\ref{app:num}).

{In practice, each step in \eqref{equ:def-I-step} corresponds to one Monte Carlo sweep (corresponding to $\tilde L^3$ insertion/deletion attempts).  The $\Delta \mu_j$ are adjusted so that one expects a typical change of $\delta n_j$ in the average number of small particles on step $j$, for a bulk system of small particles alone.  The value of $\delta n_j$ depends on the overall volume fraction and on the accuracy required: Smaller values of $\delta n$ lead to more accurate results (slower annealing during the integration of ${\cal I}$), but the computational expense is higher.  Very small $\delta n$ is required at large $\etaS$, because of significant MC rejection rates in these crowded systems.  Further details are given in the relevant sections, below.
 }

\section{Computational Details}
\label{sec:comp}

\subsection{Computation of CG potentials}
\label{app:cg-comp}

\noindent As a first application of this theory, we explain the derivation of $V_2$ and $V_3$ (and $\Delta\mu$) in the CG model.  

For the two-body potential $V_2$, consider a configuration $\CC_r$ that contains exactly two particles ($N=2$), separated by a distance $r$.  The exact two-body effective potential is (by definition) 
\beq
V_2^{\rm exact}(r) = \Phi(\CC_r) - \Phi(\CC_\infty)
\eeq
Since $\Phi(\CC_r)$ can be estimated from (\ref{equ:crooks}), this quantity can be estimated.
If one also considers the configuration $\CC_{(0)}$ which has no large particles at all, and the configuration $\CC_{(1)}$ with exactly one large particle, the exact one-body term in the CG model is
\beq
\Delta \mu^{\rm exact} = \Phi(\CC_{(0)}) - \Phi(\CC_{(1)})
\eeq
which allows $\Delta \mu$ to be estimated by (\ref{equ:crooks}).  
One may also fix $\Phi(\CC_\infty) = 2\Phi(\CC_{(1)}) - \Phi(\CC_{(0)})$.

This procedure provides point estimates of $V_2$ at equally-spaced values of $r$; a smoothed estimate of $V_2$ is obtained by fitting to a continuous function, and then tabulated for use in simulations of the CG model.  (See Fig.~\ref{fig:cgPot}(a), discussed below in 
Appendix~\ref{sec:supp-results}.)  A similar method enables computation of the three-body interaction potential $V_3$, using systems with three large particles.

So far the method is identical to~[\RefKoba].  However, two aspects of the three-body potential are different from that work.
Firstly, we set $V_3(r_{12}, r_{23}, r_{13}) = 0$ unless $\sigma_B < r_{ij} < \sigma_B+0.8\sigma_S$ for all pairs of particles.  (It is expensive to estimate this function to high accuracy, so it is convenient to set it to zero in regions of space where its value is not much larger than the numerical error.  Small errors in $V_3$ will be corrected by the two-level method in any case.)
We tabulate $V_3(x,y,z)$ for $x,y,z$ on a cubic grid with spacing $\sigS/10$, and we use linear interpolation to estimate its value for generic arguments.

The second difference from~[\RefKoba] is that we compute $V_3$ based on a deterministically chosen set of large-particle configurations (a random sample was used in~[\RefKoba]).  These samples correspond to the points of the cubic grid described above, and the symmetry of $V_3$ under interchange of all arguments is ensured by ordering the arguments by increasing size.

We require high accuracy in these free energy estimates.% these (very accurate) computations of the CG pair potential (see below), 
so we use small values for the parameter $\delta n$ that is used in the estimate of (\ref{equ:def-I-step}).   For $V_2$ we take $\delta n = 10^{-3}$ for $\etaS \leq 0.2$, also
$\delta n=5\times 10^{-4}$ for $0.2 < \etaS \leq 0.3$, and
$\delta n=6.25\times 10^{-5}$ for $0.3 < \etaS \leq 0.35$.   For computation of the three-body potential, larger systems are required (hence more expensive computations) but less accuracy is needed, so we increase $\delta n$ by a factor of 2.5.

\subsection{Computation of $\Delta\hat{p}$}
%\label{app:cg-comp}

In order to estimate $\Delta p$ using (\EqHatDp), we take $M$ representative configurations of the CG model, denoted by ${\cal C}_1,\dots,{\cal C}_M$, obtained by GCMC simulation of the CG model.
For each sample, we compute $\hat W({\cal C}_\alpha)$.  Then define a normalised reweighting factor
\beq
\hat \omega_\alpha = \frac{ \hat W({\cal C}_\alpha) }{ \frac{1}{M} \sum_{\beta=1}^M  \hat W({\cal C}_\beta)} \; .
\label{equ:omega}
\eeq
With this choice, it is shown in~[\RefKoba] that (\EqHatDp) is an appropriate estimate of $\Delta p$, 
in the sense that its mean converges for large $M$ to the true $\Delta p$, and its variance converges to zero~[\RefKoba].  We emphasise that this property holds even if the CG model is not accurate, although very large $M$ may be required in that case.

 We note that each estimate of $\hat{W}({\cal C}_\alpha)$  requires a GCMC simulation for the small particles that may take several days on a single CPU core.  
However, all the $\hat{W}$ computations are independent, allowing efficient use of high-performance (parallel) computing resources.
In practice, we make four independent estimates of the weight $\hat{W}$ for each coarse configuration; the average of these estimated weights is used as an unbiased estimate of the true weight.

For the results of the main text we take $M=1280$.  
When computing the reweighting factors $\omega_\alpha$ in the two-level method we take $\delta n = 10^{-2}$ for $\etaS \leq 0.2$, also
$\delta n=5\times 10^{-3}$ for $0.2 < \etaS \leq 0.3$, and
$\delta n=6.25\times 10^{-4}$ for $0.3 < \etaS \leq 0.35$.  (This is a suitable compromise between accuracy and computational time.)

\section{Supplementary Results}
\label{sec:supp-results}

\subsection{Coarse-grained model}

\begin{figure*}
\includegraphics[height=5.4cm]{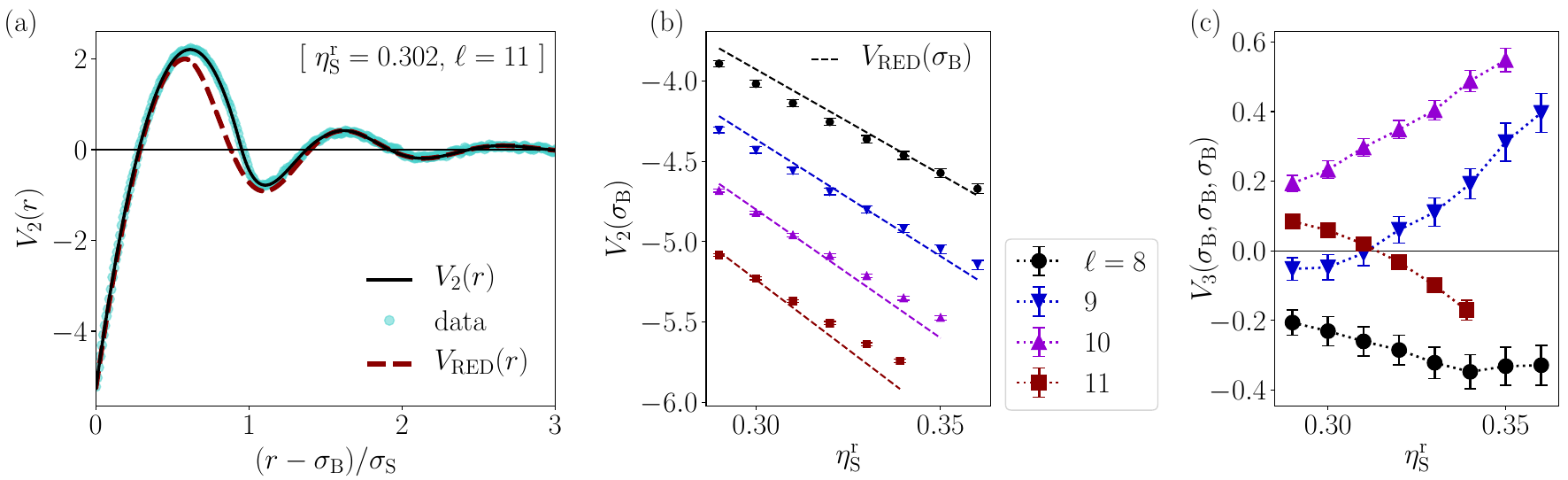}
\caption{\textbf{CG model.} (a) Tabulated two-body interaction $V_2(r)$ for parameters $\etaS=0.302$ and $\ell=11$ (close to the critical point).  This is shown together with the data from which it is estimated. The RED potential~[\RefRED] is shown for comparison, it is mostly consistent with the data, but it underestimates the repulsive part of the potential. (b) The strength of the two-body interaction is illustrated using the value of the potential $V_2$ when the particles are touching.  The strength increases (that is, the potential becomes more negative) on increasing $\eta_S^{\rm r}$ and $\ell$.  Dashed lines are a comparison with the RED potential.  (c) The strength of the three-body interaction is illustrated using the value of $V_3$ when all three particles are touching.  The dependence on the model parameters more complicated in this case, as discussed in the main text.  Dotted lines are guides to the eye.}
\label{fig:cgPot}
\end{figure*}

\noindent Fig.~\ref{fig:cgPot} illustrates the behavior of the effective interactions in the CG model.  We give a brief description of its main properties.

Fig.~\ref{fig:cgPot}(a) shows the two-body effective interaction, which has the form of a depletion potential.  
There is a strong effective attraction between the particles, whose range is comparable with $\sigS$.  Also, the layering of the small particles around the large ones means that the potential has oscillations, with both attractive and repulsive parts.   We show results for parameters close to the critical point of the model, which are compared with the potential proposed by Roth, Evans and Dietrich (RED)~[\RefRED].  As previously noted in~[\RefAshton], the RED potential is close to the true $V_2$, but there are significant differences in the repulsive parts of these potentials. 
The error bars on $V_2$ are no larger than symbol sizes, hence the depletion potential is accurate.

Fig.~\ref{fig:cgPot}(b) shows how the strength of the depletion interaction depends on the size ratio $\ell$ and on $\etaS$.  This is quantified by the value of the depletion potential at contact.  As expected, the potential gets stronger as $\ell$ and $\etaS$ increase.

By contrast, Fig.~\ref{fig:cgPot}(c) indicates the strength of the three-body potential, for the specific case where all three particles are touching each other.  (The strong two-body attraction means that this arrangement is the most common, so it is suitable for illustrative purposes.)  The three-body potential is smaller in absolute value than $V_2$, and it may be either attractive ($V_3<0$) or repulsive ($V_3>0$).  Note also that there is no clear trend for the dependence on $\etaS$ and $\ell$: the potential may increase or decrease.  

\subsection{Three-body interactions and definition of $\phi_{\rm in}$}

As discussed in the main text, the dependence of $V_3$ on model parameters is related to the packing of the small particles around the large ones.
The quantity $\phi_{\rm in}$ is defined by fixing three large particles in mutual contact (as in Fig.~\FigPhi) and simulating the small particles in the grand canonical ensemble.  Let $n_\triangle$ be average number of small particles within the shaded grey area of Fig.~\FigPhi\ (specifically, in a three-dimensional region that extends above and below the plane of the Figure by a distance $\delta z/2$ in each direction).  The area of the shaded region is $A_\triangle=(2\sqrt{3}-\pi)\sigB^2/8$ and 
\beq
\phi_{\rm in} = \frac{ \sigS^3 }{ \delta z A_\triangle } n_\triangle  
\eeq 
is the number density in the relevant volume (in units of $\sigS^{-3}$).
This quantity depends on the small-particle volume fraction, the comparison in Fig.~\FigPhi\ is at $\etaS =0.32$ and we take $\delta z \sim \sigS$.  For very large $\ell$ then $\phi_{\rm in}$ tends to the bulk number density but its behaviour for moderate $\ell$ is subtle, because of the complexity of the underlying  sphere packings.

\begin{figure*}
\includegraphics[width=18cm]{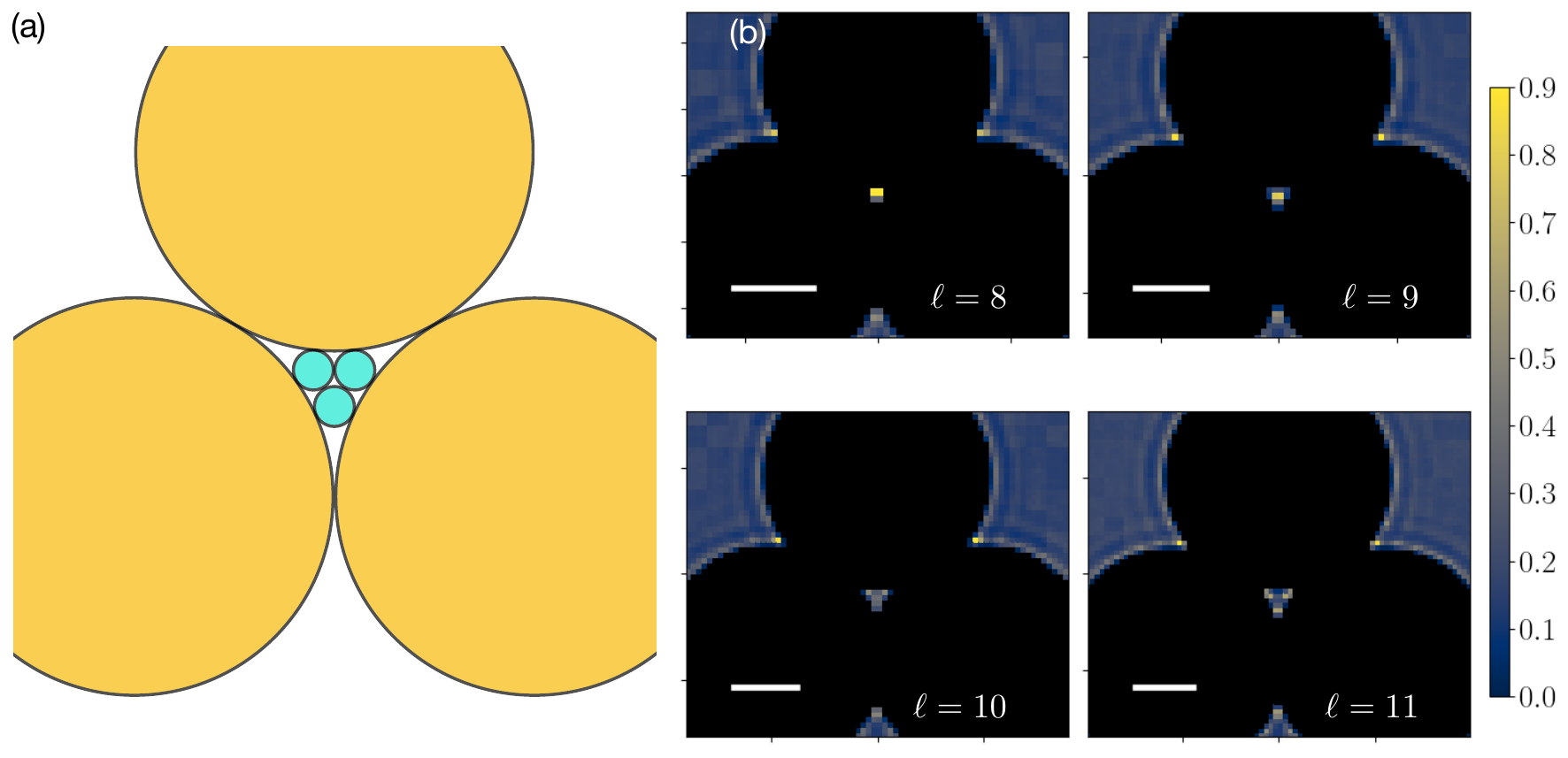}
\caption{(a) Illustration of the packing of small particles in the vicinity of three large ones.   This (planar) configuration is possible for size ratios $\ell \geq 5+\sqrt{24} = 9.90$.  (b) Local density of small particles in the vicinity of three (fixed) large ones for $\etaS=0.32$ and $\ell=8,9,10,11$.  In the case $\ell=11$, three particles fit the gap, visible as three local peaks in the density.  
Scale bars are $3\sigS$.}
\label{fig:maps}
\end{figure*}

Fig.~\ref{fig:maps} presents additional information to allow the behavior of $V_3$ to be rationalized.  It shows the density of small particles in the vicinity of three large ones, which have fixed positions, all touching each other.  For $\ell=11$, three particles can fit into the (approximately) triangular region between the particles, while for $\ell\leq 10$, this does not occur.  (Exactly at $\ell=10$, three small particles can just fit in the planar arrangement of Fig.~\ref{fig:maps} but their positions are tightly constrained and the associated phase-space volume is extremely small.)  As a result, the packing for $\ell=11$ is much more efficient than for $\ell=10$, and the corresponding $V_3$ is smaller.  By contrast, for $\ell=8$, putting a single small particle into this region corresponds to a relatively efficient packing and a smaller $V_3$, at least compared with $\ell=9,10$.

These three-body effects have many subtle  features.  For the purposes of this work, two aspects are important.  First, the potential at contact has values that are smaller than unity, but these are certainly not negligible contributions to the energy.   Second, the sign of the interaction (and its dependence on $\etaS$) has a non-trivial dependence on $\ell$.  Specifically, the three-body effect for $\ell=10$ is significantly repulsive (and increasingly so at large $\etaS$), while the corresponding effect for $\ell=11$ is weakly repulsive for $\etaS\approx 0.3$ but becomes attractive at larger $\etaS$.

\subsection{\normalsize Discussion and further results for FG model}

\begin{figure}
\centering
\includegraphics[width=8.5cm]{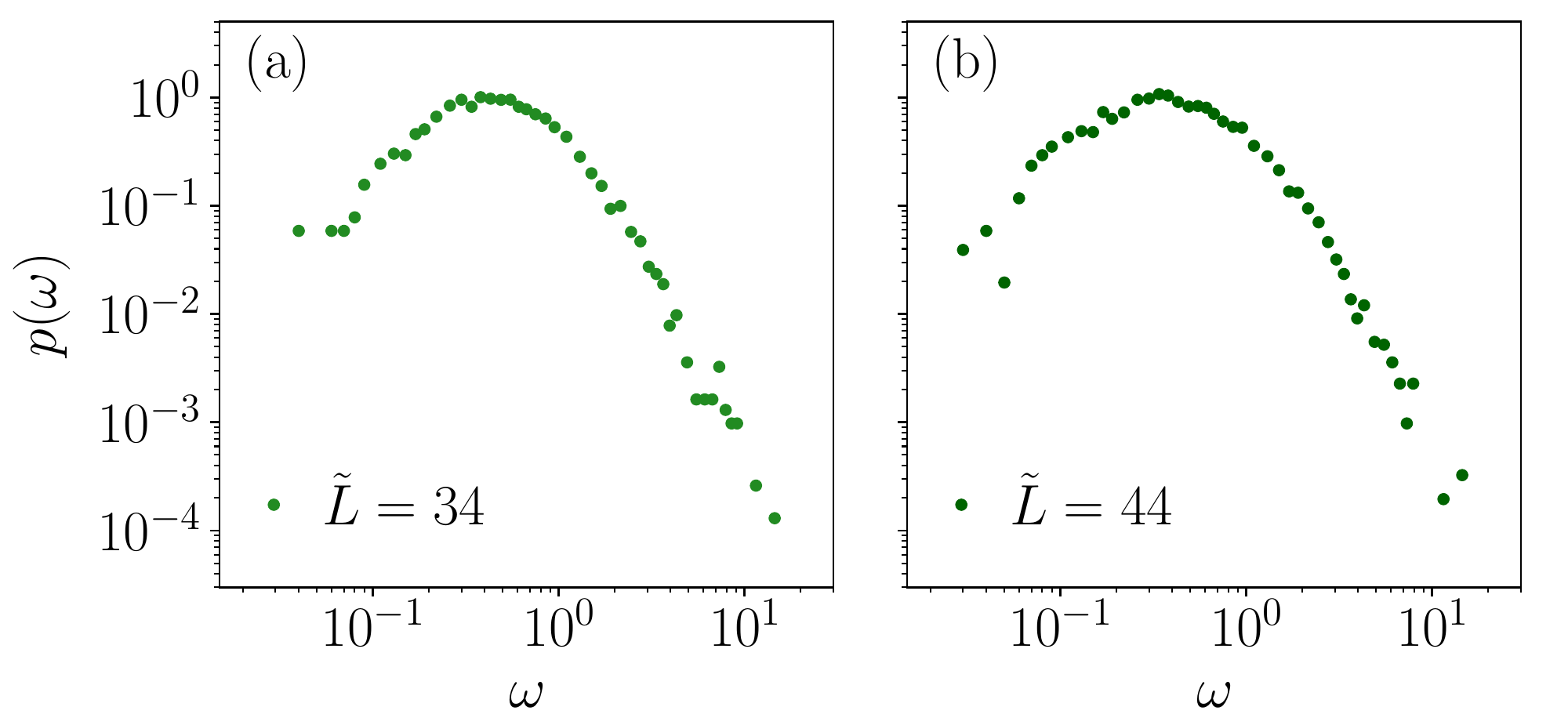}
\caption{Distributions of the reweighting factors $\omega_\alpha$ for the FG models of Fig.~\FigHistos. %\ref{fig:histos}.  
The important feature is that large reweighting factors (above $\omega=10$ for example) are rare.}
\label{fig:omega}
\end{figure}

\noindent \textbf{Reweighting factors and accuracy of CG model}:
\\
As a consistency check between the CG and FG models, Fig.~\ref{fig:omega} shows the distribution of $\hat\omega_\alpha$.  This distribution has $\langle \hat\omega \rangle=1$ by construction, but its variance has a significant impact on the results of the two-level method.  In particular, if the CG model is not accurate (or the Jarzysnki integration has large variance) then there will be some configurations with very large $\omega_\alpha$: these tend to dominate the estimate 
Eq.~\EqHatDp,
%\eqref{equ:hat-dp}, 
resulting in a large statistical uncertainty in $\Delta \hat p$.  (An example of this effect was shown in~[\RefKoba].)  

Both histograms in Fig.~\ref{fig:omega} show a few samples with $\hat\omega\approx10$, which have some impact on the FG results in Fig.~\FigHistos.  
In particular, the data for $\tilde L=44$ are somewhat scattered in that Figure.  
Still, the kernel density estimate for $p(N)$ reduces the uncertainty by averaging over several values of $N$, and appears to yield reliable estimates.

Recalling that a perfect coarse-grained model would have energy function $E_{\rm CG}^{\rm ex}(\CC) = \Phi(\CC)$ (up to an additive constant), it is useful to define the Kullback-Leibler divergence between the Boltzmann distributions of our CG model and the exact one, which is
\beq
D_{\rm KL}^{\rm CG} = \sum_N \int  d\bm{R}_1\dots d\bm{R}_N \, p_{\rm CG}(\CC) \log [ Z / W(\CC)  ] 
\label{equ:KL-ZW}
\eeq
with $W(\CC)$ as in (\ref{equ:def-W}).
This $D_{\rm KL}^{\rm CG}$ is non-negative and measures how different is the CG model from the exact one.   It is zero if (and only if) the CG model is exact.  This may be observed by writing it in the form 
\beq
D_{\rm KL}^{\rm CG} = \langle E_{\rm CG}^{\rm ex}(\CC) - E_{\rm CG}(\CC) \rangle_{\rm CG} + \log \langle {\rm e}^{E_{\rm CG}(\CC) - E_{\rm CG}^{\rm ex}(\CC)} \rangle_{\rm CG}
\label{equ:KL}
\eeq
which shows that it can be interpreted as the average coarse-graining error in $E_{\rm CG}$.  

In a free-energy perturbation theory computation~[\RefZwanzig], this quantity could be computed.  In the method used here, the $W(\CC)$ are not available but we do have their (unbiased) estimates $\hat{W}(\CC)$.   Consider the quantity
\beq
\hat{D} = -\frac{1}{M} \sum_\alpha \log \hat{\omega}_\alpha 
\eeq
with $\hat\omega$ as in (\ref{equ:omega}), and recall that the configurations $\CC_\alpha$ are representative samples from the CG model.
Since $\hat{W}(\CC)$ is an unbiased estimate of $W(\CC)$, we have $\langle \hat{W}(\CC_\alpha) \rangle_{\rm J} = W(\CC_\alpha)$ where $\langle \cdot \rangle_{\rm J}$ is the expectation value with respect to the stochastic computation of $\hat W$, see also [\RefKoba].
By (\ref{equ:omega}) we have
 \beq
 \langle \hat{D} \rangle_{\rm J} =  -\frac{1}{M} \sum_\alpha  \left\langle \log \hat{W}(\CC_\alpha) \right\rangle_{\rm J} + \left\langle \log \frac{1}{M} \sum_\alpha \hat{W}(\CC_\alpha) \right\rangle_{\rm J}
 \eeq
For large $M$ then $\frac{1}{M} \sum_\alpha \hat{W}(\CC_\alpha) \approx \langle W \rangle_{\rm CG} = Z$ (because  the $\CC_\alpha$ are representative CG configurations).  Also, Jensen's inequality means that $ \langle \log \hat{W}(\CC_\alpha) \rangle_{\rm J}  \leq  \log\langle  \hat{W}(\CC_\alpha) \rangle_{\rm J} = \log \hat{W}(\CC_\alpha)$.  Using these facts we obtain
  \beq
 \langle \hat{D} \rangle_{\rm J} \gtrsim   -\frac{1}{M} \sum_\alpha \log \hat{W}(\CC_\alpha) +  \log Z
 \eeq
 Finally using again that the $\CC_\alpha$ are representative coarse configurations we have 
   \beq
 \langle \hat{D} \rangle_{\rm J} \gtrsim   \langle \log [Z/\hat{W}(\CC)]  \rangle_{\rm CG} 
 \eeq
The right hand side is the KL divergence as in (\ref{equ:KL-ZW}) so we finally obtain
 \beq
 D_{\rm KL}^{\rm CG} \lesssim  \langle \hat{D} \rangle_{\rm J}
 \eeq
 That is, the computable quantity $\hat{D}$ is an estimated upper bound for the error $ D_{\rm KL}^{\rm CG}$ of the CG model.
 
 From the distributions of Fig.~\ref{fig:omega}, we estimate $\hat{D} \approx 0.32$ for $\tilde{L} = 34$ and $\hat{D} \approx 0.37$ for $\tilde{L} = 44$.  Hence, the error of the (total) energy of a configuration in the CG model is less than $0.4$ (in units of $k_{\rm B}T$, relative to an exact coarse-grained model).   Since these are total energies for systems with significant numbers of particles, this indicates that the two- and three-body interactions are indeed accurate.
 
\vspace{6pt}\noindent \textbf{The role of crystallisation}:
\\
An additional advantage of the two-level method arises because the critical point in binary hard sphere systems is metastable to crystallization.  In our study, crystallisation of the large particles was sometimes observed during simulation of the CG model.  Since this CG model is relatively easy to simulate, we take the simple approach of discarding those CG simulation runs where crystallization occurs; this still allows generation of sufficient data, at manageable cost.  In any method that requires full simulation of the FG model, crystallisation events are disastrous because they require large quantities of costly data to be discarded.  (Such effects might be mitigated by automated methods for avoiding crystallisation, but this is not simple to achieve, without biasing the sampling of fluid states.)

\vspace{6pt}\noindent \textbf{Finite size effects and field mixing in $p(N)$}:
\\
The systems considered in this work are relatively small, compared to the diameter of the large particles, which does affect the results.  However, the finite-size scaling theory of the critical point is well-developed~[\RefWild], which allows these effects  to be rationalised.
In particular, one sees from Fig.~\FigHistos(a,d) that the probability $p(0)$ is not completely negligible, so the system may contain no large particles at all.  The universal form is relevant for large $N$ -- it does not account for the fact that this number is an integer, nor that it must be non-negative.  Hence one cannot expect an exact match to this form in small systems.   So-called field-mixing effects arising from the lack of symmetry between the fluid phases~[\RefWild] can also lead to asymmetry in $p(N)$, resulting in deviations from the (symmetric) scaling form for finite-sized systems.  
Larger systems would allow a more detailed analysis of these effects, as well as estimation of critical exponents.  However, given the various types of corrections to scaling that should be expected, the close agreement observed here between the numerical data and the universal form is remarkable, and represents strong evidence for a de-mixing critical point.

\vspace{6pt}\noindent \textbf{Extended law of corresponding states}:
\\
Noro and Frenkel~[\RefNoro] proposed that critical points for systems with short-ranged attractive (two-body) potentials can be estimated by a criterion based on the reduced second virial coefficient, which in this context is $B_2^* = (3/\sigB^3) \int_0^\infty [1-{\rm e}^{-V_2(r)} ] r^2 dr$. (The factor of $3$ is included so that $B_2^*=1$ for a hard sphere potential.)  They defined 
\beq
\tau = \frac{1}{4(1-B_2^*)} 
\eeq
so that small positive $\tau$ corresponds to strong attractive interactions.
For short-ranged attractive systems, they found that critical points generically occur for $\tau\approx 0.1$.
 For adhesive hard sphere (AHS) models (corresponding to very short-ranged attractive attractions), it was later estimated~\RefMiller that $\tau\approx 0.113$ at criticality.  This can be interpreted as an (extended) law of corresponding states~[\RefNoro].
 
For the potentials studied here, we find for the critical parameters $\ell=11$ and $\etaS\approx 0.30$ that $\tau=0.11$.   
The three-body effect is weak at this state point:
if we revert to a two-body CG model with the same parameters, the system is close to criticality.
 For $\ell=10$,  Fig.~\FigHistosSigTen(b) indicates that the two-body CG system is critical for $\etaS$ slightly below $0.348$, corresponding again to $\tau\simeq 0.1$, similar to~[\RefAshton].  These results are consistent with the extended law of corresponding states.
 
 \vspace{18pt}\noindent \textbf{Behavior for very large $\ell$}:
\\
We offer a few comments on the limit of large $\ell$, corresponding to very extreme size ratio.  
 This limit $\ell\to\infty$ is quite subtle~[\RefDijkstraPRE].  It is convenient to fix $\sigB$ and take $\sigS\to0$.  This can be done in three different ways: (i) keeping the concentration of small particles constant~\RefVega; (ii) keeping the volume fraction of small particles constant~\RefStell; (iii) keeping the second virial coefficient $B_2^*$ constant, for the effective interactions~\RefBaxter.  
 
 There is obviously no demixing in case (i)~[\RefVega], and crystallisation tends to dominate in case (ii)~[\RefDijkstraPRE].  As noted in \RefMiller, the relevant case for fluid-fluid demixing is (iii).  In this case one expects [\RefDijkstraPRE] that  $\etaS \sim (1/\ell) \log \ell$, which tends to zero $\ell\to\infty$.  For very small $\etaS$, interactions among the small particles can be neglected and we expect the system to behave similarly to an Asakura-Oosawa model, with a short-ranged two-body attraction, and negligible three-body and higher contributions.  The qualitative behavior that we find for $\ell=11$ is consistent with this physical picture: that two-body interactions dominate for very large $\ell$ and fluid-fluid demixing should occur.  However, it is not clear how large $\ell$ should be in general, for three-body interactions to have a negligible effect.

\end{appendix}

\bibliography{multi}
\end{document}